\documentclass{pazhastl}

\usepackage{graphicx}
\usepackage{natbib}

\usepackage{color}

\def\smfigurewocap#1#2#3{
  \begin{minipage}{1.0\columnwidth}
    \begin{minipage}{0.049\columnwidth}
      \rotatebox{90}{\phantom{0000}#3}
    \end{minipage}
    \begin{minipage}{0.9\columnwidth}
      \includegraphics[bb=58 188 556 678,width=0.97\columnwidth]{#1}
      \centerline{#2}
    \end{minipage}
    
    \vskip 3pt
    ~
  \end{minipage}
}

\begin{document}

\journalinfo{2008}{34}{10}{653}[663] 

\title{Optical Identifications of Five INTEGRAL Hard X-ray Sources\\ in the
  Galactic Plane Region}

\author{I.~F.~Bikmaev\address{1,2}\email{ilfan.bikmaev@ksu.ru},
  R.~A.~Burenin\address{3}, 
  M.~G.~Revnivtsev\address{3,4},
  S.~Yu.~Sazonov\address{3,4},
  R.~A.~Sunyaev\address{3,4},
  M.~N.~Pavlinsky\address{3},
  N.~A.~Sakhibullin\address{1,2}
  \addresstext{1}{Kazan State University, ul. Kremlevskaya 18, Kazan, Russia} 
  \addresstext{2}{Academy of Sciences of Tatarstan, ul. Baumana, 20, Kazan,
    Russia}
  \addresstext{3}{Space Research Institute, ul. Profsoyuznaya 84/32, Moscow,
    Russia}
  \addresstext{4}{Max-Planck Institut fur Astrophysik,
    Karl-Schwarzschild-Str.\ 1, Garching, Germany}
}

\shortauthor{Bikmaev et al.} 
 
\shorttitle{Optical Identifications of Five INTEGRAL Hard X-ray Sources}

\submitted{\today}

\begin{abstract}
  The results of optical identifications of five hard X-ray sources in the
  Galactic plane region from the INTEGRAL all-sky survey are presented. The
  X-ray data on one source (IGR\,J20216+4359) are published for the first
  time. The optical observations were performed with 1.5-m RTT-150 telescope
  (TUBITAK National Observatory, Antalya, Turkey) and 6-m BTA telescope
  (Special Astrophysical Observatory, Nizhny Arkhyz, Russia). A blazar,
  three Seyfert galaxies, and a high-mass X-ray binary are among the
  identified sources.
  
\end{abstract}

\section*{INTRODUCTION}

The INTEGRAL all-sky survey carried out during the last few years (see,
e.g., \citealt{krivonos07}) provides an opportunity to study the nearby
active galactic nuclei (AGNs), accreting white dwarfs, high-mass and
low-mass X-ray binaries, symbiotic stars, etc. The advantage of INTEGRAL
energy range (17--60~keV) is that it allows to be almost completely free
from the selection effect related to photoabsorption of X-ray emission both
near the observed X-ray source and on the line of sight in the Galactic
interstellar medium.

A considerable number of hitherto unknown hard X-ray sources have been
discovered during this survey. Our group performs optical identifications of
these sources in the northern sky with the Russianâ Turkish 1.5-m RTT-150
telescope \citep{bikmaev06a,bikmaev06b,burenin08}. In this paper, we present
the results of optical identifications of a set of sources from the INTEGRAL
all-sky survey located near the Galactic plane.

\section*{OBSERVATIONS}

As usual, for our observations we chose a number of northern-sky objects
($\delta>-30^\circ$), for which accurate positions in the sky were known
from observations with the X-ray telescopes onboard ROSAT, Chandra, and
SWIFT observatories. The Chandra data for several sources were obtained in
frames of projects proposed by our group \citep{sazonov05,sazonov08}. We
retrieved all necessary additional publicly available X-ray data from the
HEASARC archive\footnote{http://heasarc.nasa.gov/}.

The optical observations of the sources were carried out with RTT150
telescope in the spring and summer of 2007, using two instruments --- the
CCD photometer based on the thermoelectrically cooled Andor CCD and the low-
and medium-resolution spectrometer
TFOSC\footnote{http://astroa.physics.metu.edu.tr/tug/tfosc.html}.  

For spectroscopy we used low resolution grism \#15, which give the highest
efficiency and the most wide spectral range (3300--9000~\AA). In this setup
spectral resolution is $\approx15$~\AA\ (FWHM). In addition, some of the
sources were observed with 6-m BTA telescope in the fall of 2007, using
spectrometer SCORPIO \citep{scorpio}. The data were reduced using the
standard IRAF\footnote{http://iraf.noao.edu} and
DECH\footnote{http://www.gazinur.com/Download.html} \citep{dech} software
packages.

\section*{RESULTS OF OBSERVATIONS}

The list of studied sources and their classifications are given in the table
~\ref{tab:1}.  The coordinates of the optical objects are given at epoch
J2000, according to the astrometric solutions in RTT150 direct images, which
were obtained relative to the USNO-B1.0 catalog \citep{monet03}. The
magnitudes were measured using direct images obtained with RTT150. The
photometric calibration was done using the observations of standard stars
from \cite{landolt}.  Below, the X-ray data and the results of optical
observations for each source are discussed in more detail.

\begin{table*}
  \caption{The list of identified sources}
  \label{tab:1}
  \begin{center}
    \begin{tabular}{lcccclcc}
      \hline
      \hline
      Name&\multicolumn{2}{c}{$\alpha$, $\delta$ (J2000)}&$R_c$&$z$&Type$^{1}$\\
      \hline
      RX\,J0137.7+5814 & 01 37 50.45 & +59 14 11.6 &17.63&  ? & BL Lac \\
      IGR\,J20216+4359 & 20 21 49.04 & +44 00 39.4 &19.14&  $0.017$\phantom{$73$}& Sy2\\
      IGR\,J21343+4738 & 21 34 20.37 & +47 38 00.4 &13.79&  $-$& HMXB\\
      IGR\,J23206+6431 & 23 20 36.58 & +64 30 45.2 &19.41& 0.07173& Sy1\\
      IGR\,J23523+5844 & 23 52 22.00 & +58 45 32.7 &18.62&  $0.1620$\phantom{$3$}& Sy2\\
      
      \hline
    \end{tabular}
    \medskip
    
    \begin{minipage}{0.7\linewidth}
      \footnotesize
      
      $^{1}$ Sy1, Sy2 --- type 1 and 2 Seyfert galaxies; BL Lac --- BL
      Lacertae object; HMXB --- High Mass X-Ray Binary;

    \end{minipage}
  \end{center}
\end{table*}

\begin{figure}
  \centering
  
  \includegraphics[width=0.8\columnwidth]{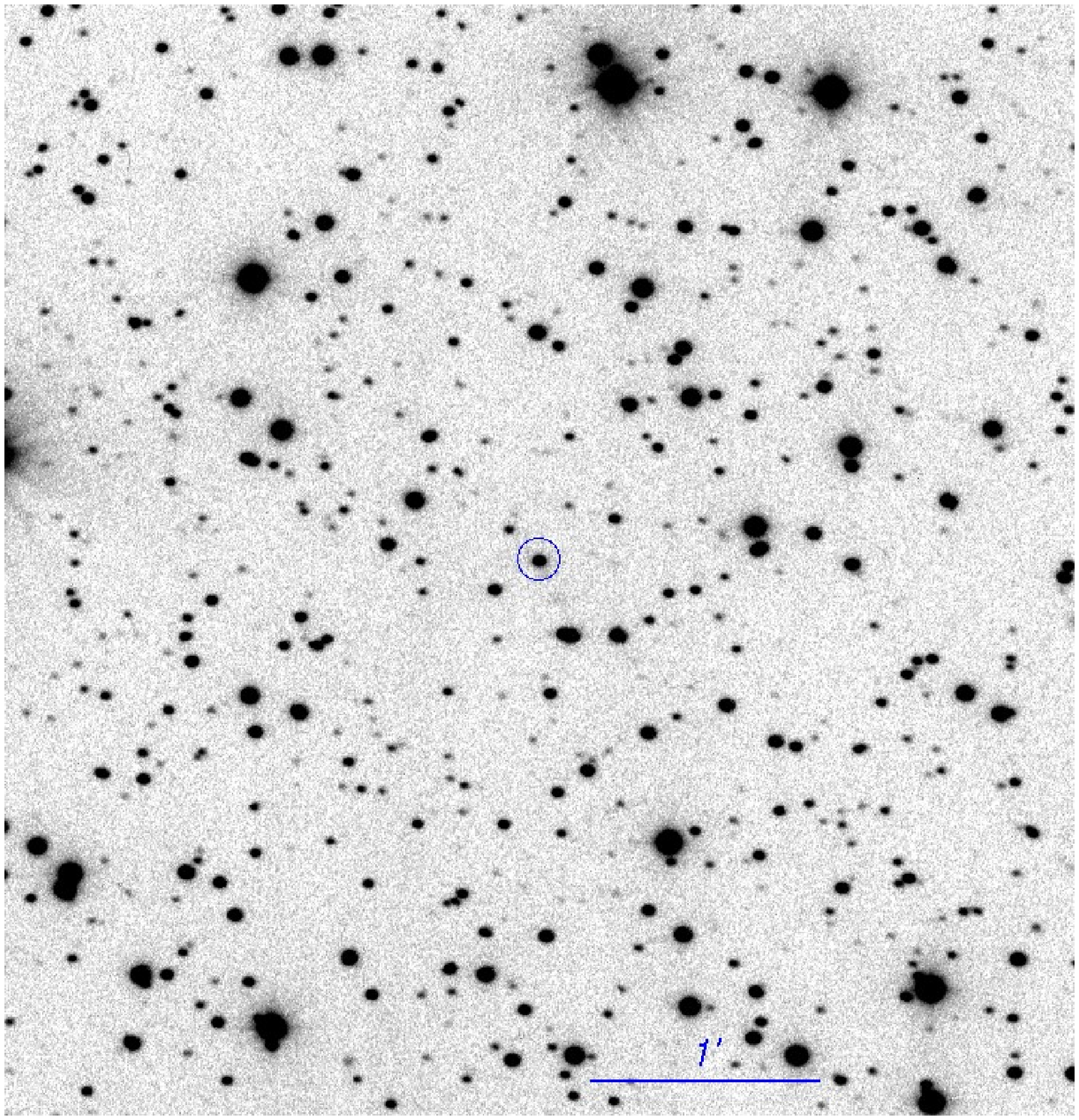}
  \bigskip
  
  \smfigurewocap{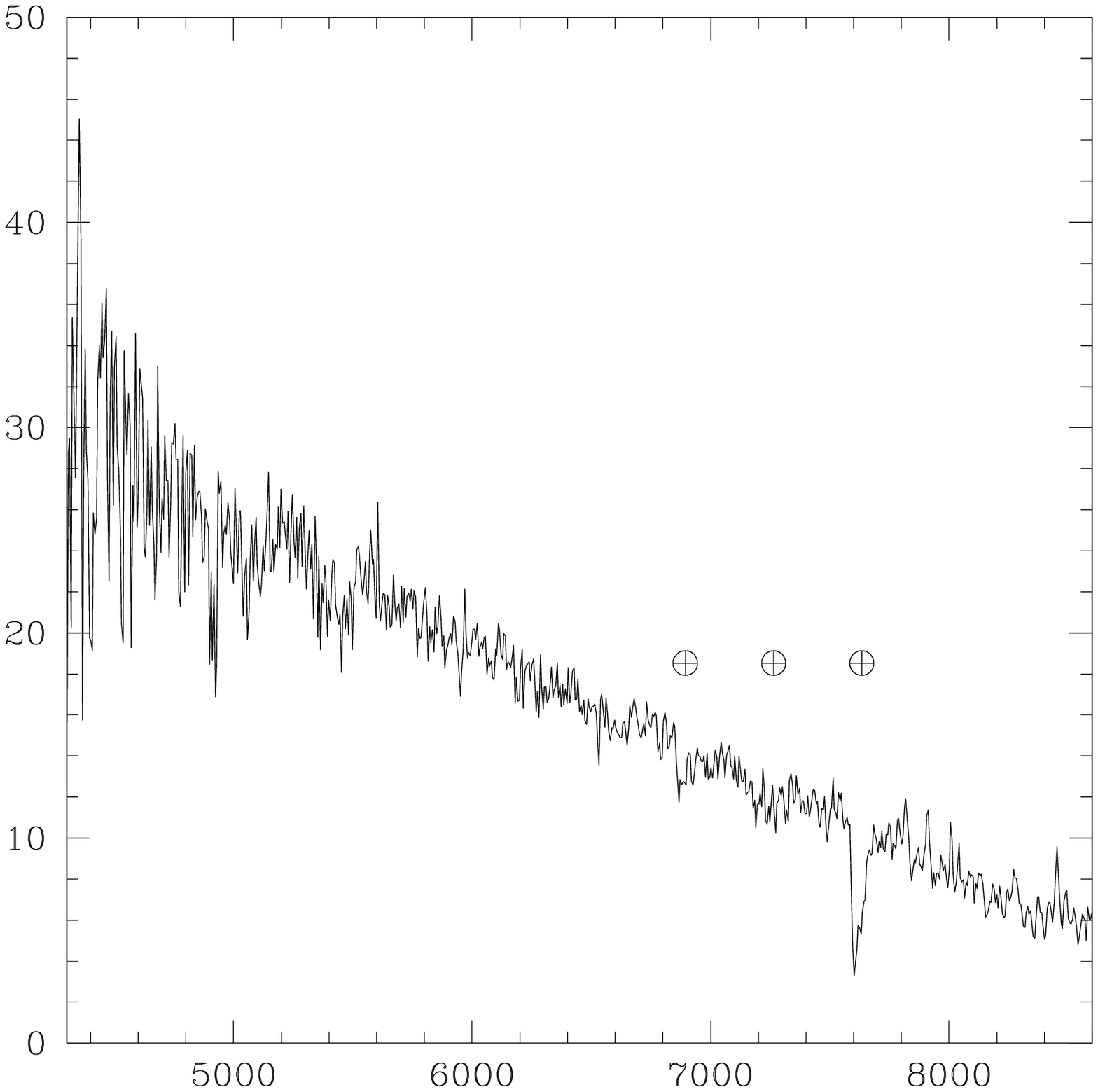}{$\lambda$, \AA}{$F_\lambda,
    \times 10^{-16}$~ÜÒÇ~Ó$^{-1}$~ÓÍ$^{-2}$~\AA$^{-1}$}
  
  \caption{Upper panel --- optical image of the field around the source
    RX\,J0137.7+5814 from RTT150 Rc-band observations. The circle marks the
    $\approx6\arcsec$ error circle of the position of the radio source
    87GB\,013433.2+575900. Lower panel --- combined optical spectrum of
    RX\,J0137.7+5814 obtained from RTT150 and BTA observations corrected for
    the Galactic extinction $E(B-V)=0.85$.}
  \label{sp_rxj0137}
\end{figure}

\subsection*{RX\,J0137.7+5814}

The position of the X-ray ROSAT source RX\,J0137.7+5814 \citep{voges99} and
the corresponding hard X-ray INTEGRAL source \citep{krivonos07} coincides,
within the error limits, with the bright radio source 87GB\,013433.2+575900
whose position is known with an accuracy of $\approx6-10\arcsec$.  Only one
star with magnitude $R<18$ lies within this $10\arcsec$ circle
(Fig.~\ref{sp_rxj0137}, upper panel).

The spectra of this source were obtained with RTT150 during several nights.
In addition, the spectrum of this object was also taken with the 6-m BTA
telescope using the SCORPIO spectrometer (Afanasiev and Moiseev 2005).
Because of poor weather conditions during these observations we failed to
perform measurements with the required high signal-to-noise ratio.
Nevertheless, the data were of comparable quality to those from RTT150.

The combined RTT150 and BTA spectrum of the optical object corrected for the
Galactic extinction $E(B-V)=0.85$ is shown in Fig.~\ref{sp_rxj0137} (lower
panel). Telluric absorption bands (near Ë 6900, 7200, and 7600~\AA) and,
probably, some unidentified absorption lines (e.g., the line at 4914~\AA)
are clearly seen in this spectrum. However, the spectrum exhibits no
detectable stellar absorption line and there are no emission lines that are
observed in AGN spectra. Together with the presence of the intense radio
emission, this suggests that RX\,J0137.7+5814 is a blazar or, more
precisely, a BL Lac object. More sensitive optical observations of the
source are needed in order to measure its redshift.

\begin{figure}
  \centering
  \includegraphics[width=0.8\columnwidth]{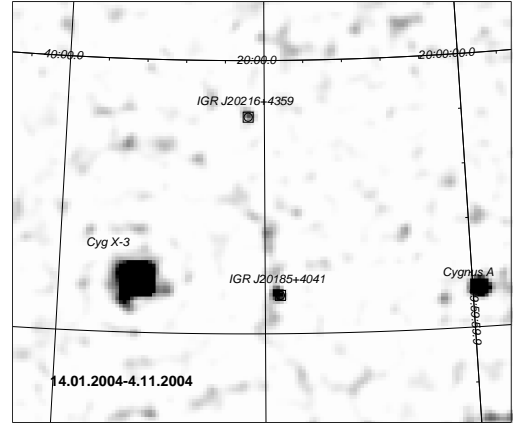}
  \caption{The IBIS/INTEGRAL image of the field around the source
    IGR\,J20216+4359 in the 17--60~keV energy range.}
  \label{igrj20218_integral}
\end{figure}
 
\begin{figure}
  \centering
  \includegraphics[width=0.8\columnwidth]{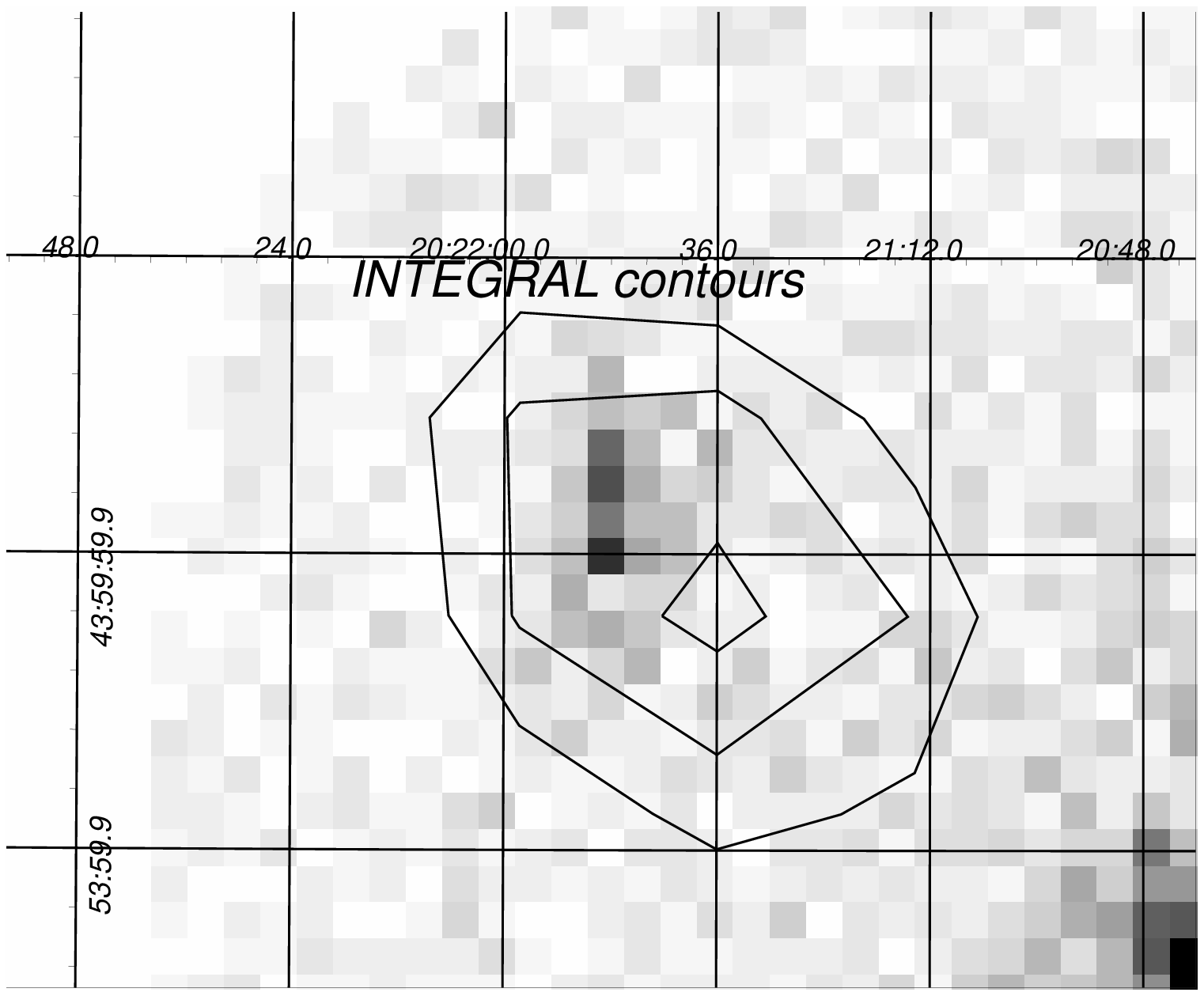}
  
  \includegraphics[width=0.8\columnwidth]{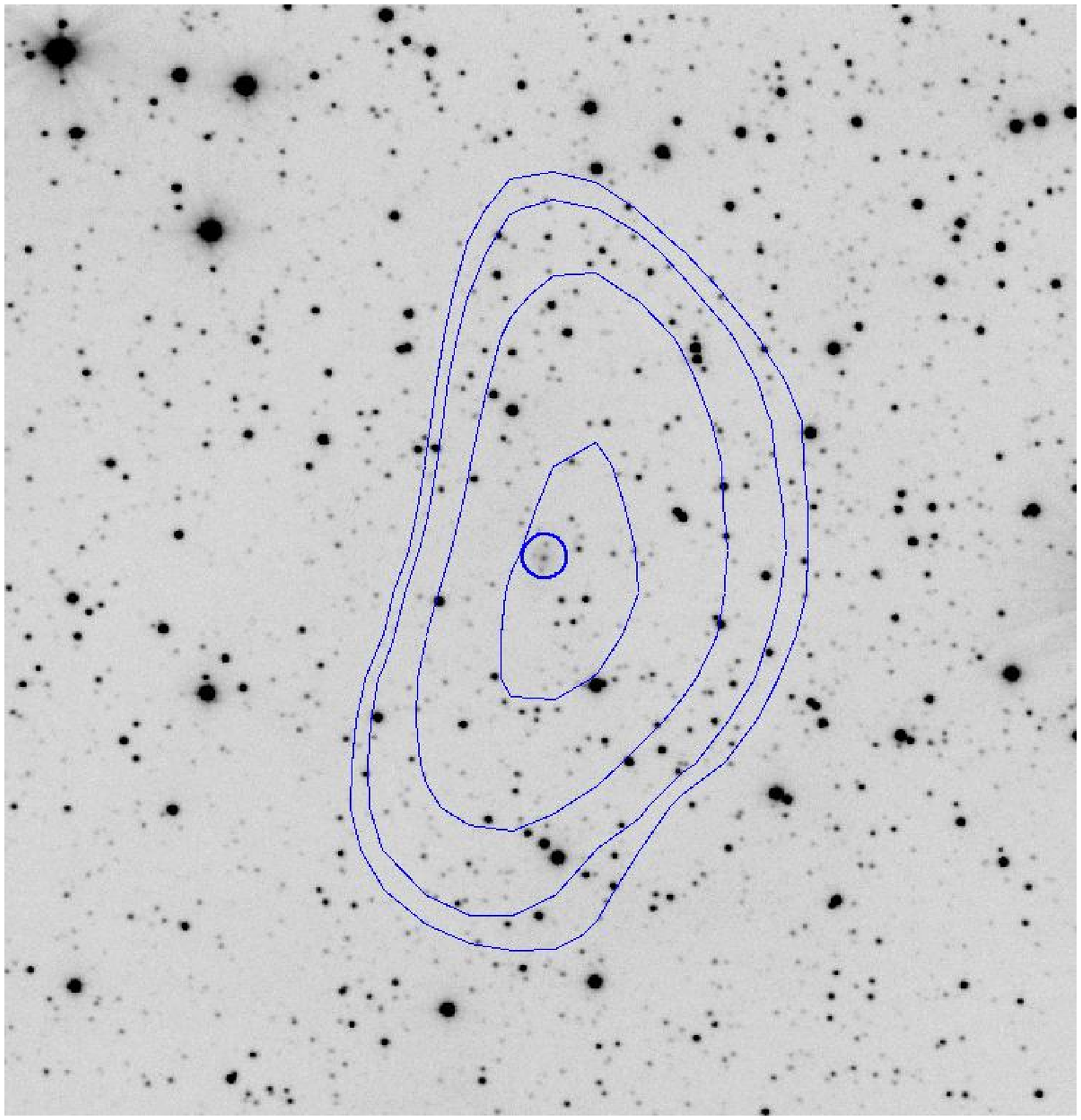}
  \caption{Upper panel --- the ASCA image of the field around the source
    IGR\,J20216+4359 in the energy range 4--10~keV obtained during the
    observations on June 10, 1993. The contours show the 4.0, 4.5, and
    5$\sigma$ IBIS/INTEGRAL flux levels. Lower panel --- optical image of
    the field around IGR\,J20216+4359 from RTT150 observations. The contours
    indicate the ASCA position of the X-ray source. The circle marks the
    galaxy whose active nucleus is an X-ray source.}
\label{igrj20218_asca_dss}
\end{figure}

\begin{figure}
  \centering
  \smfigurewocap{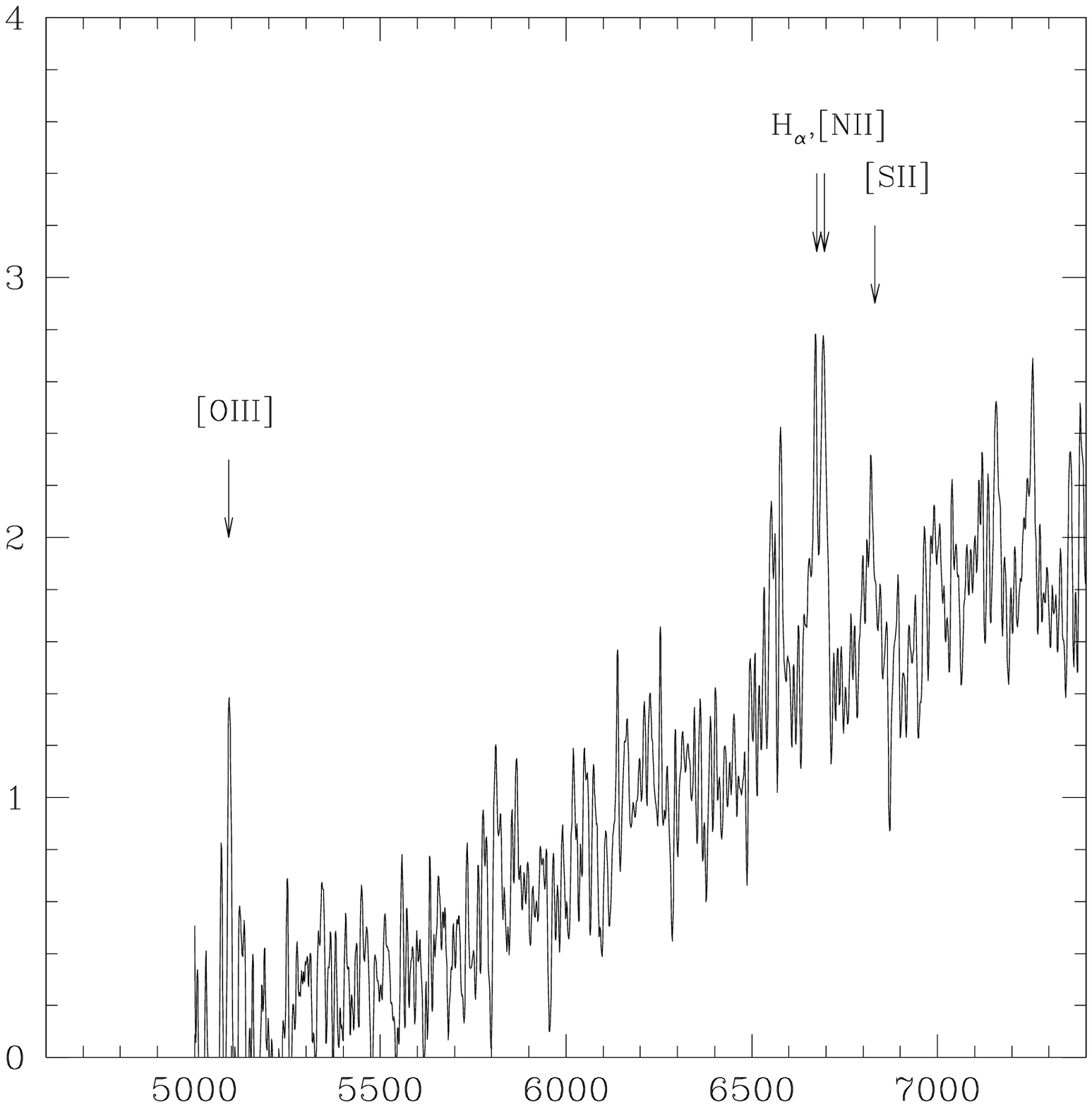}{$\lambda$, \AA}{$F_\lambda,
    \times 10^{-16}$~ÜÒÇ~Ó$^{-1}$~ÓÍ$^{-2}$~\AA$^{-1}$}
  \caption{The spectrum of the optical counterpart of the source
    IGR\,J20216+4359, obtained with RTT150 telescope, not corrected for the
    Galactic extinction.}
  \label{igrj20218_sp}
\end{figure}

\subsection*{IGR\,J20216+4359}

The hard X-ray source IGR\,J20216+4359 was discovered in an incomplete set
of INTEGRAL observations of the Cygnus region (observations from January 14,
2004, to November 4, 2004; orbits 153--251). In this series of observations,
the source was detected at a high confidence level ($\approx5.5\,\sigma$,
Fig.~\ref{igrj20218_integral}). Its coordinates are $\alpha$=21 21.8,
$\delta$=+43 59 (J2000), the positional accuracy is $\approx3\arcmin$. The
hard X-ray flux (17--60~keV) from the source was $\approx1.1$~mCrab, which
corresponds to the energy flux of $\approx1.6\times
10^{-11}$~erg~s$^{-1}$~cm$^{-2}$ for the powerlaw spectrum with a photon
index $2$. In the complete set of observations of the Cygnus region, the
source is detected at a confidence level of only $\approx3\,\sigma$ which
corresponds to a flux of $0.6\pm0.2$~mCrab.

The field around IGR\,J20216+4359 was observed by ASCA observatory on June
10, 1993. The source was detected at a statistically significant level in
these observations, which allow to improve its coordinates: $\alpha,
\delta$: 20 21 48.1 +44 00 32 (J2000, the accuracy is $\approx20\arcsec$,
Fig.~\ref{igrj20218_asca_dss}) and to measure its spectrum in the standard
X-ray range 0.8--10~keV. The source turned out to be strongly absorbed,
i.e., the equivalent photoabsorption column density measured from the shape
of its X-ray spectrum is $n_H L=(13\pm2)\times10^{22}$~cm$^{-2}$ (the photon
index was fixed at $\Gamma=1.7$ due to poor statistics). This value is
considerably higher than that in the Galactic interstellar medium, which is
$\approx10^{22}$~cm$^{-2}$ \citep{dickey90}. The strong internal X-ray
photoabsorption is typical for type 2 Seyfert galaxies.

The Seyfert galaxy was found in the error box of IGR\,J20216+4359 using
optical RTT150 observations. This galaxy is marked by the circle in
Fig.~\ref{igrj20218_asca_dss}. Its spectrum (Fig.~\ref{igrj20218_sp})
exhibits forbidden [OIII] and [NII] emission lines, which indicate the high
AGN activity \citep[see, e.g.,][]{baldwin81,kauffmann03}. The absence of
broad $H\alpha$ and $H\beta$²lines in the AGN spectrum confirms the
classification of this AGN as a type 2 Seyfert galaxy. The redshift of the
galaxy measured using narrow [OIII] and [NII] lines is $z=0.017$.

\begin{figure}
  \includegraphics[width=\columnwidth,bb=27 277 566 567]
  {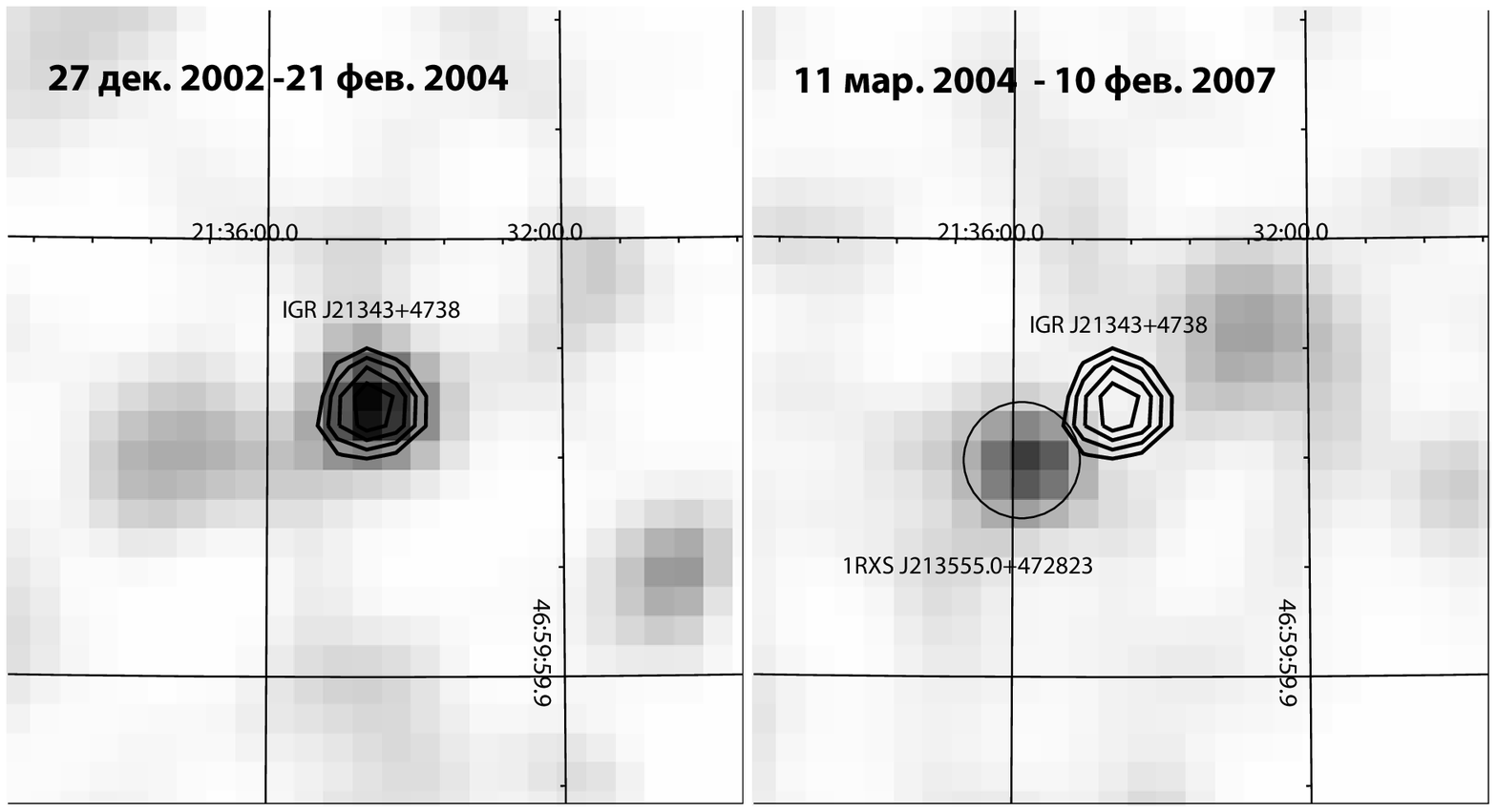}
  \caption{INTEGRAL 17--60~keV hard X-ray images of the field around
    IGR\,J21343+4738 from December 27, 2002, to February 21, 2004 (left),
    and from March 11, 2004, to February 10, 2007 (right). The source
    IGR\,J21343+4738 was clearly detected in the first observing period and
    was below the threshold in the second one. The AGN
    1RXS\,J213555.0+472823 is detected near IGR\,J21343+4738 during the
    second observing period. The contours in the images denote the regions
    of equal statistical significance of the flux in the image obtained in
    the first observing period starting from $3.5\sigma$ and with
    $0.5\sigma$ steps.}
  \label{igrj21343_integral_images}
\end{figure}

\subsection*{IGR\,J21343+4738}

The hard X-ray source IGR\,J21343+4738 was discovered during deep
observations of the Galactic-plane region in Cygnus
\citep{krivonos07,bird07}. More detailed studies of the sources behavior
showed it to be variable. Fig.~\ref{igrj21343_integral_images} shows the
IBIS/INTEGRAL images of the sky field around IGR\,J21343+4738 in different
observing periods. IGR\,J21343+4738 is detected at a statistically
significant level in the IBIS/INTEGRAL observational data only in the series
of observations from December 27, 2002, to February 21, 2004
(Fig.~\ref{igrj21343_integral_images}, left panel). The mean 17--60~keV flux
from the source was $1.6\pm0.3$~mCrab in this series of observations, which
corresponds to the flux
$\approx(2.3\pm0.4)\times10^{-11}$~erg~s$^{-1}$~cm$^{-2}$.

\begin{figure}
  \centering
  \includegraphics[width=0.7\columnwidth]{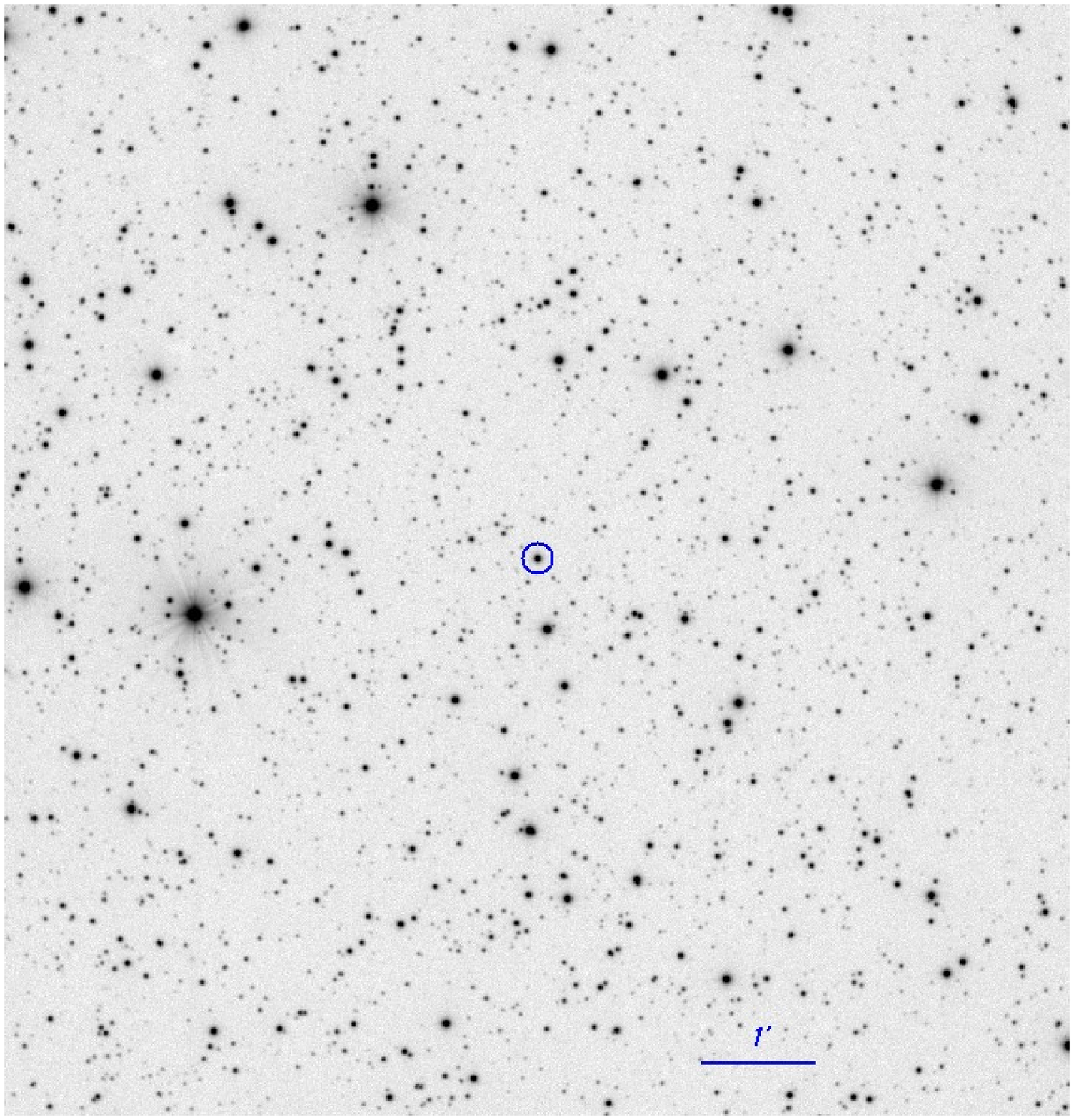}
  \caption{The direct image of the field of the source IGR\,J21343+4738 in
    filter $R$, obtained with RTT150 telescope.}
  \label{igrj21343r}
\end{figure}

\begin{figure}
  \centering
  \includegraphics[width=0.8\columnwidth,bb=62 30 700
  550,clip]{IGR21343Fluxes.PDF.eps}
  \includegraphics[width=0.8\columnwidth]{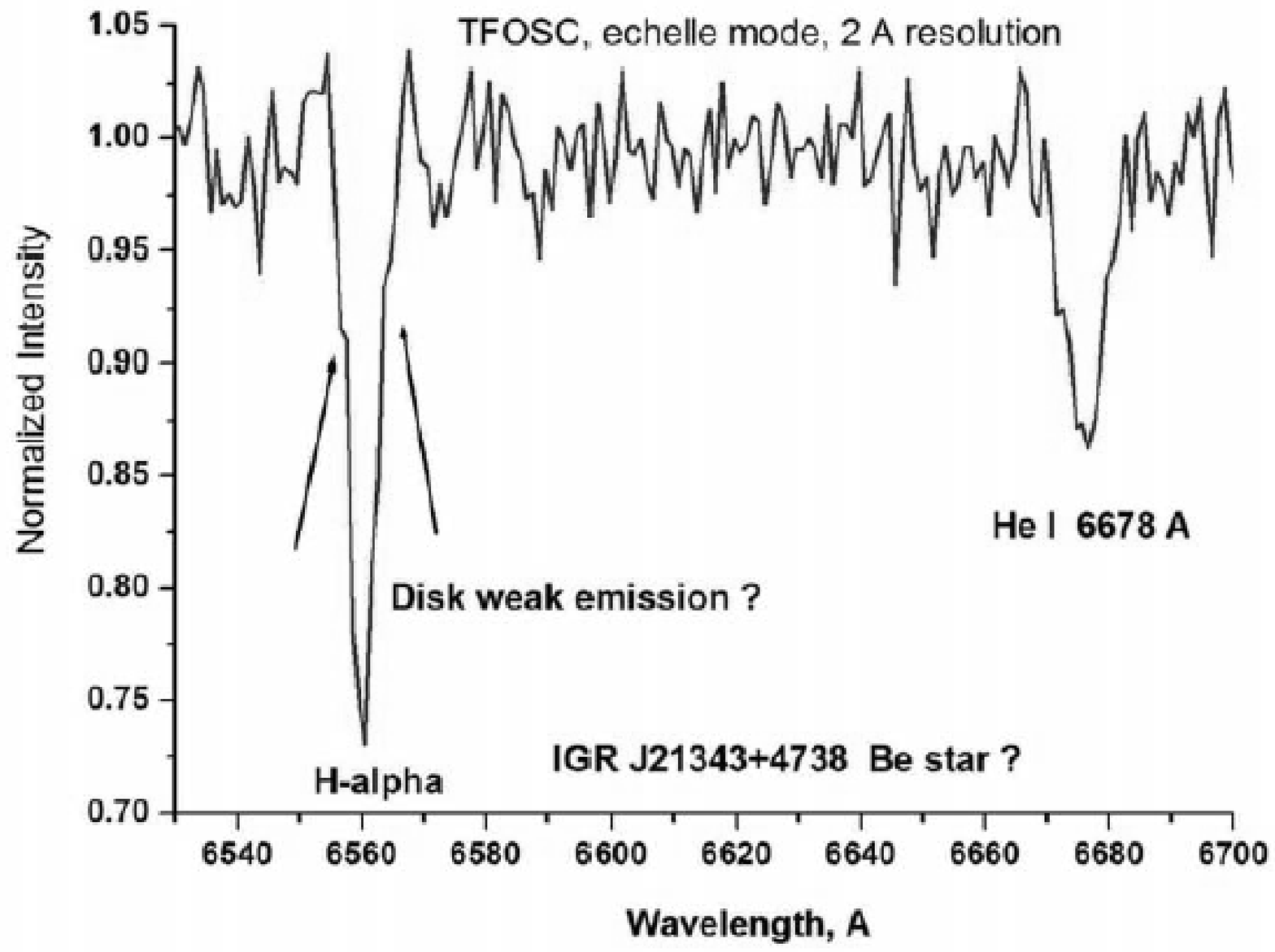}
  \caption{Upper panel --- spectrum of the optical object identified with
    the source IGR\,J21343+4738. Lower panel --- higher-resolution spectrum
    of the source near the $H\alpha$ line. The arrows indicate the possible
    contribution from weak emission lines of the equatorial disk of a
    massive optical star.}
  \label{igrj21343}
\end{figure}

The source is not detected in the map of this field averaged over the period
from March 11, 2004, to February 10, 2007
(Fig.~\ref{igrj21343_integral_images}, right panel). In this series of
observations, the exposure time was much longer than that in the first one
and the upper limit on the source 17--60~keV flux was 0.5~mCrab ($2\sigma$),
suggesting its transient nature. Due to the higher sensitivity, the AGN
RX\,J2135.9+4728 \citep{burenin08} is detected in this image near the
location where the source IGR\,J21343+4738 was previously found. This AGN is
located at a distance of $\approx15\arcmin$ from IGR\,J21343+4738, which is
considerably larger than both the IBIS localization accuracy and its angular
resolution. Using the other bright sources in the IBIS field of view, we
checked that the astrometric errors in these observations are small.

The field around IGR\,J21343+4738 was observed by the Chandra observatory
on December 18, 2006 \citep{sazonov08}. Based on the INTEGRAL observations,
one may expect the sources brightness to drop significantly compared to that
in the first observing period. However, the high sensitivity of the Chandra
observatory allows to detecd a weak hard X-ray source in the error region of
the hard X-ray source IGR\,J21343+4738 that can be unambiguously associated
with the optical object with the following coordinates $\alpha, \delta$: 21
34 20.37 +47 38 00.4 (J2000). The finding chart for this field is shown in
Fig.~\ref{igrj21343r}.

The spectrum of this object (Fig.~\ref{igrj21343}) shows signatures of B3
star. In particular, in addition to the overall shape of the spectrum, the
fairly intense HI and HeI absorption lines that are too strong for A type
stars and and the absence of strong HeII absorption lines typical of O stars
point to the B spectral type. This implies that the X-ray source is most
likely a high-mass X-ray binary. In this case, the transient nature of this
source in X rays is not unusual, especially in view of the recent discovery
of a large number of the so-called fast X-ray transients in highmass X-ray
binaries \citep[see, e.g.,][]{chaty07}.

In our optical observations we detected $H\alpha$ line in absorption, not in
emission, as it is usually observed in high-mass X-ray binaries. It may be
related to the long-period evolution of the equatorial disk wind from the
optical companion similar to what is observed for some Be systems
\citep[see, e.g.,][]{norton91}. The higher resolution echelle spectrum
obtained with the TFOSC spectrometer (resolution $\approx2$~\AA,
Fig.~\ref{igrj21343}, lower panel) showed that $H\alpha$ line is more narrow
than HeI,6678~\AA, which may indicate the presence of a weak double-peaked
emission from the equatorial disk of the optical star.

\begin{figure}
  \centering
  \includegraphics[width=0.7\columnwidth]{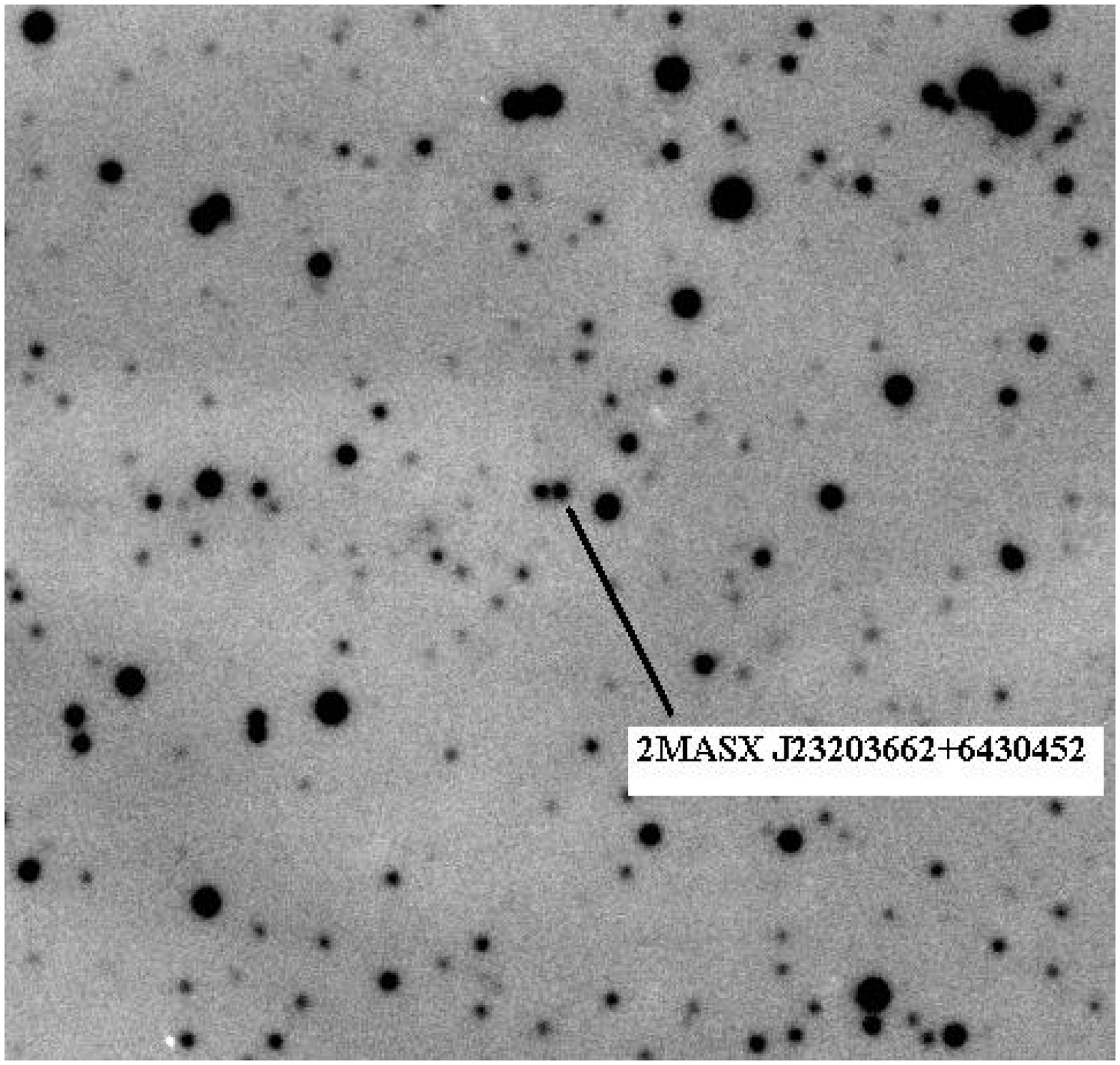}
  \includegraphics[width=0.9\columnwidth]{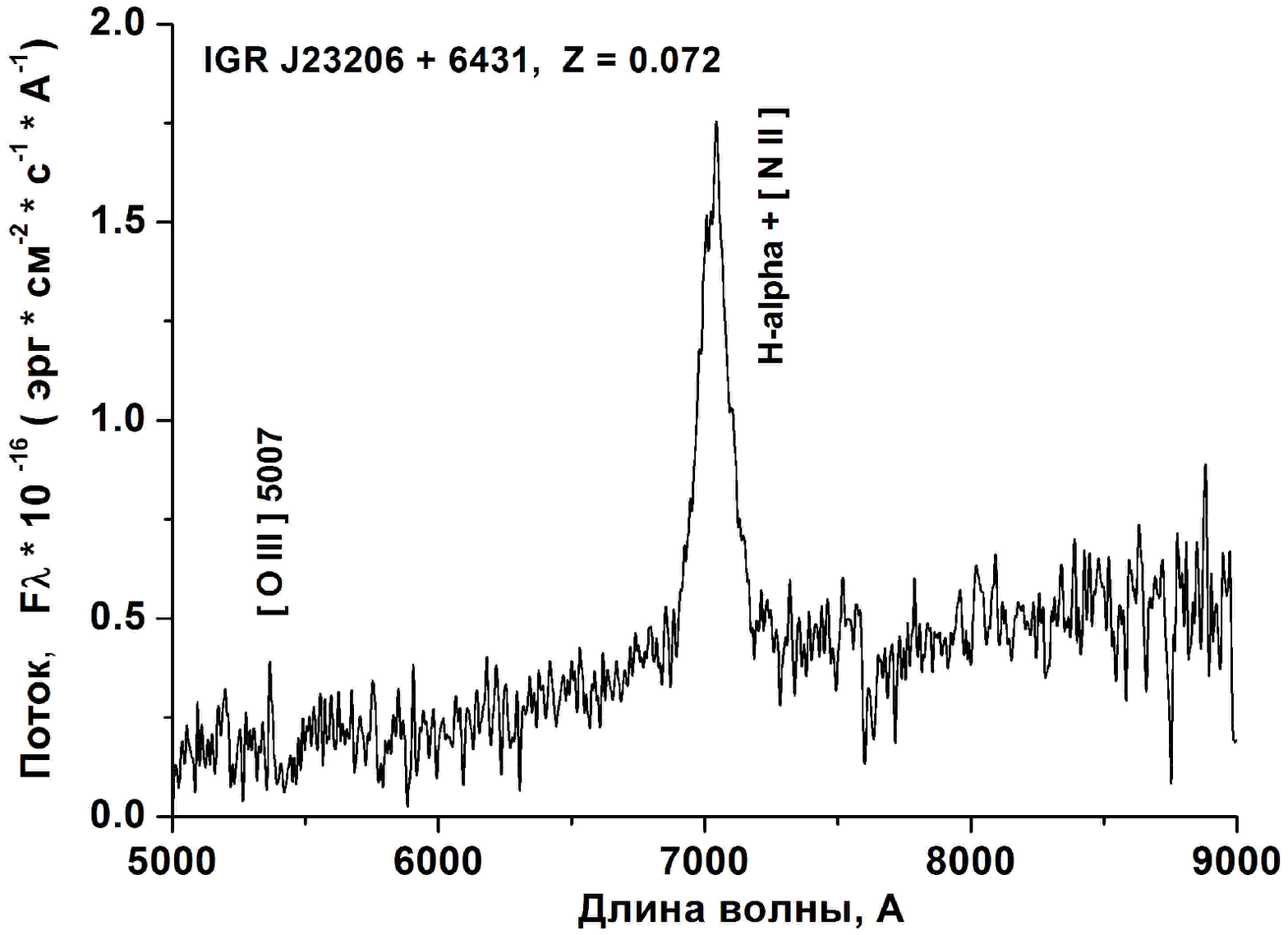}
  \caption{Optical R-band image of the IGR\,J23206+6431 field (upper panel)
    and optical spectrum of the source from RTT150 observations (lower
    panel).}
  \label{igrj2320}
\end{figure}

\subsection*{IGR\,J23206+6431}

The hard X-ray source was discovered in the IBIS/INTEGRAL images of the
Galactic plane region, which were obtained after the catalog of the all-sky
survey \citep{krivonos07} was published. The time-averaged 17--60~keV flux
from the source is $0.6\pm0.1$~mCrab or $\approx8.7\times
10^{-12}$~erg~s$^{-1}$~cm$^{-2}$. This source was observed with the X-ray
telescope onboard SWIFT observatory, which allowed to unambiguously identify
it with the galaxy 2MASX\,J23203662+6430452.

Figure~\ref{igrj2320} shows the direct image of the sources field (upper
panel) and its optical spectrum (lower panel) not corrected for the Galactic
extinction obtained with RTT150. The optical spectrum of this object
exhibits redshifted broad $H\alpha$ and narrow [OIII],5007 lines. Thus, this
source is a type 1 Seyfert galaxy. Its redshift derived from the [OIII] line
is $z=0.07173$. Preliminary information about the optical identification of
this source was immediately published in astronomical circular
\citep{atel1363}.

\begin{figure}
  \centering
  \includegraphics[width=0.8\columnwidth]{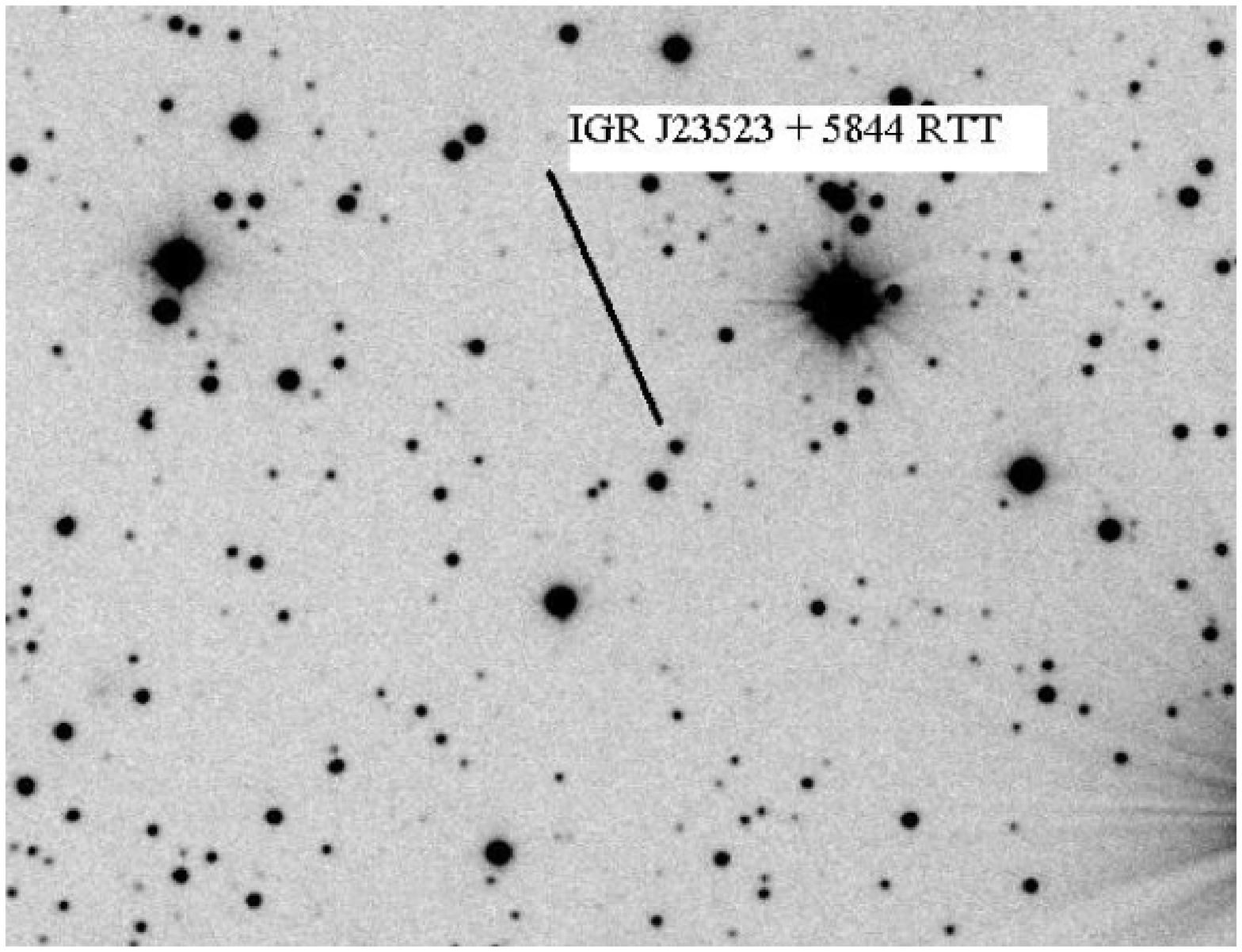}
  \includegraphics[height=0.9\columnwidth,bb=132 37 538 640,clip,angle=90]
  {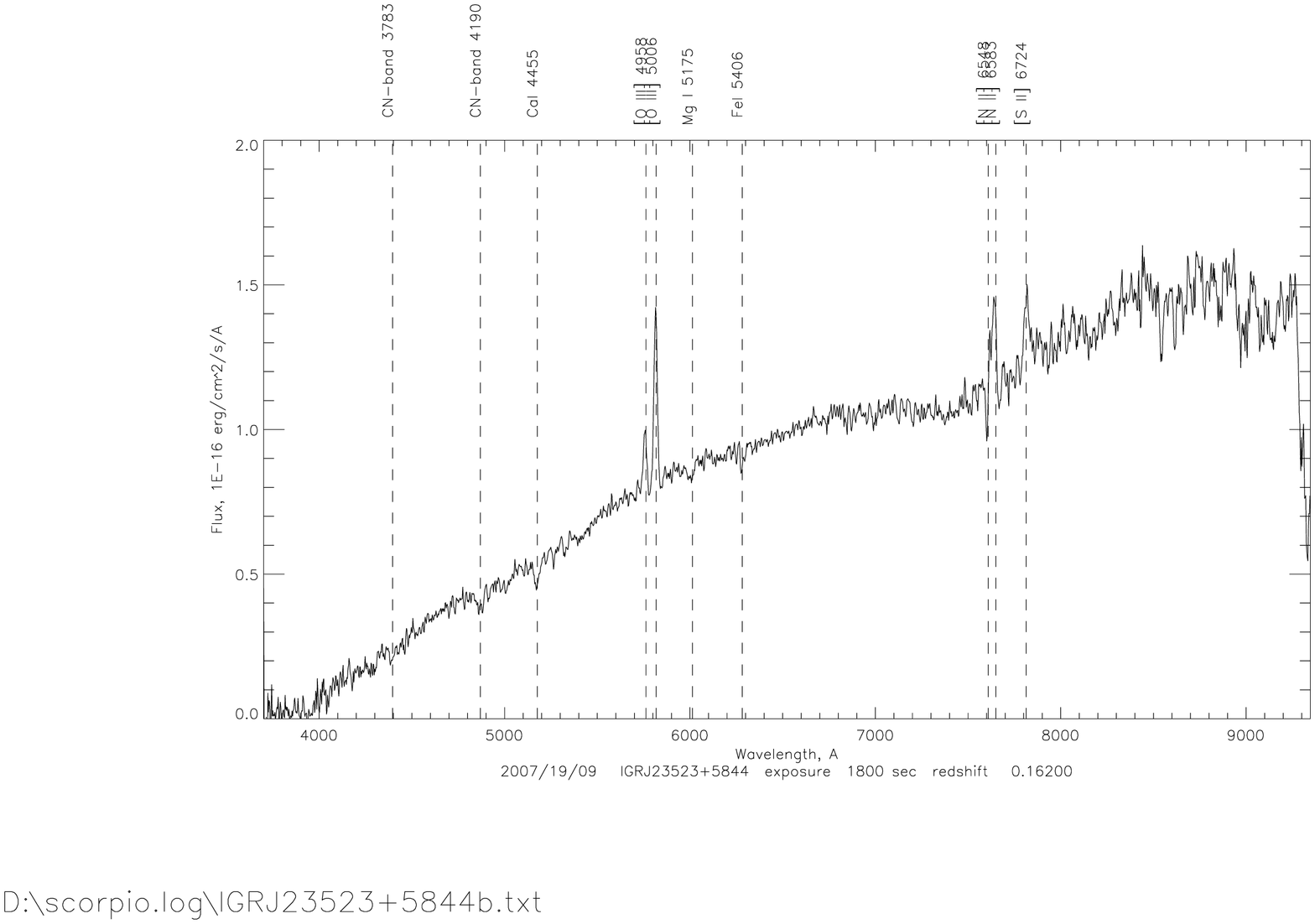}
  \caption{Upper panel --- optical image of the field around the source
    IGR\,J23523+5844 obtained with RTT150 telescope. Lower panel --- optical
    spectrum of the source obtained with BTA, not corrected for the Galactic
    extinction.}
  \label{igrj23523}
\end{figure}

\subsection*{IGR\,J23523+5844}

The hard X-ray source was observed by Chandra observatory on January 14,
2007 \citep{sazonov08}. The X-ray absorption column density estimated from
the Chandra X-ray spectrum is $n_H L=(3.7\pm0.5)\times 10^{22}$~cm$^{-2}$,
which is much higher than the absorption column density in our Galaxy
\citep{dickey90}. This is a distinctive feature of type 2 Seyfert galaxies
in X-rays.

The accurate position of the X-ray source allowed to identify this source
with the optical object whose coordinates are given in the
Table~\ref{tab:1}. The finding chart for the field near this object is shown
in Fig.~\ref{igrj23523} (upper panel) and its optical spectrum obtained with
6-m BTA telescope using spectrometer SCORPIO \citep{scorpio} is presented in
Fig.~\ref{igrj23523} (lower panel). The spectrum of the optical object
exhibits Ca, Mg, Fe, and other absorption lines as well as intense narrow
[OIII] 4959,5007, [SII] 6717,6731, and probably [NII] 6548,6583 forbidden
emission lines. Here, the flux ratio of the [OIII] 5007 and $H\beta$² lines
is definitely larger than 10, implying that the object can be identified as
Seyfert galaxy \citep[e.g.,][]{baldwin81,kauffmann03}. The absence of an
intense broad $H\beta$ line suggests that this can be a type 2 Seyfert
galaxy.

The redshift is $z=0.1620$, the $H\alpha$ and [NII] 6548,6583 lines fall
into the atmospheric 7600 A absorption band and their observations are
complicated by the subtraction of a complex sky background. However, we can
state that the spectral feature that remains near the 7600~\AA\ band after
the correction for the atmospheric absorption is more red than $H\alpha$
line and, most likely, is the [NII] 6583 forbidden line. The spectrum also
exhibits no signs of the narrow $H\beta$ and [OII] 3727 lines which are
usually observed in the spectra of type 2 Seyfert galaxies. For example, the
lower limit on the flux ratio of the [OIII] 5007 and $H\beta$ lines is
$\approx20$ here, while the maximum value of this ratio for optically
selected AGNs is about 15 \citep[see,
e.g.,][]{baldwin81,veilleux87,kauffmann03}. Probably it can be explaned by
strong absorption of the narrow emission lines regions.

\begin{figure}
  \centering
  \includegraphics[width=0.9\columnwidth]{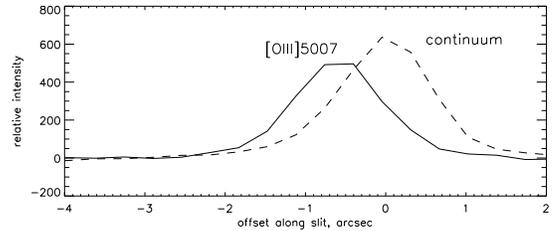}
  \caption{Intensities of the continuum and [OIII] line emission from the
    AGN IGR\,J23523+5844 along the slit of SCORPIO spectrometer.}
  \label{igrj23523oiii}
\end{figure}

We note that the [OIII] line emission in this AGN is spatially shifted from
the continuum emission. This is clearly seen from the intensity distribution
of the emission along the slit of the spectrometer SCORPIO shown in
Fig.~\ref{igrj23523oiii}. The difference in [OIII] line and continuum
surface brightness distributions is observed in some nearby AGNs, for
example, in Markarian 34 or Markarian 78, and reflects the fact that the
emission in narrow forbidden lines originates at distances of 1--2 kpc from
the central black hole \citep[e.g.,][]{1988ApJ...334..104H}. In addition,
in the case of IGR\,J23523+5844, this agrees with the suggestion that the
central regions near the AGN are strongly absorbed (see above) and the
observed emission in narrow forbidden lines originate only at considerable
distances from the AGN.

Thus, all our data taken as a whole suggest that IGR\,J23523+5844 is most
likely a type 2 Seyfert galaxy, although its optical spectrum may have some
peculiarities related to strong absorption of the central narrow emission
line regions. In their paper recently published in preprints,
\cite{masetti08} also identified this source as a probable type 2 Seyfert
galaxy.

\section*{CONCLUSIONS}

In this work we presented information on the optical identifications of five
hard X-ray sources from the INTEGRAL all-sky survey located in the
Galactic plane region. The X-ray data on one INTEGRAL source,
IGR\,J20216+4359, are published for the first time. A blazar
(RX\,J0137.7+5814), three Seyfert galaxies (IGR\,J20216+4359,
IGR\,J23206+6431, and IGR\,J23523+5844), and a high-mass X-ray binary
(IGR\,J21343+4738) are among the identified sources.

In this paper we concentrated on the objects near the Galactic plane, i.e.,
in the region that is traditionally avoided by observers in optical band
because of strong Galactic absorption and high stellar density. We see, that
the INTEGRAL data allow to discover new, previously unknown nearby AGNs that
would be very difficult to found in optical. As expected, among the hard
X-ray sources in the Galactic plane there is also a large number of Galactic
sources, mostly high-mass X-ray binaries.

\acknowledgements

We would like to thank V.~L.~Afanasiev for help with the optical
observations with 6-m BTA telescope and for a useful discussion of the
observational results. We also would like to thank A.~I.~Galeev,
R.~Ya.~Zhuchkov (Kazan State University), and I.~M.~Khamitov (TUBITAK
National Observatory) for the help with the optical observations at the
RTT150 telescope. This work was supported by Russian Foundation for Basic
Research (project nos.\ 07-02-01004 and 07-0201051), by Scientific School
Program (project nos.\ NSh-4224.2008.2 and NSh-1100.2006.2), and by Programs
P-04 and OFN-17 of Russian Academy of Sciences.

\end{document}